\documentclass{mem}

\usepackage{natbib}
\usepackage{txfonts}
\usepackage{balance}
\usepackage{graphicx}
\usepackage[a4paper]{hyperref}
\idline{00}{59}

\begin{document}

\title{Barred disks in dense environments}

\subtitle{Insights from the Abell 901/902 clusters with STAGES}

\author{ I. \,Marinova\inst{1}, S. \, Jogee\inst{1},
  A. \,Heiderman\inst{1}, F. \,D. \,Barazza\inst{3},
  M. \,E. \,Gray\inst{2}, M. \,Barden\inst{9}, C. \,Wolf\inst{14},
  C. \,Y. \,Peng\inst{5,6}, D. \,Bacon\inst{7}, M. \,Balogh\inst{8},
  E. \,F. \,Bell\inst{4}, A. \,B\"ohm\inst{10},
  J. \,A. \,R. \,Caldwell\inst{16}, B. \,H\"au\ss ler\inst{2},
  C. \,Heymans\inst{12}, K. \,Jahnke\inst{4}, E. \,van Kampen\inst{9},
  K. \,Lane\inst{2}, D. \,H. \,McIntosh\inst{13,17},
  K. \,Meisenheimer\inst{4}, S. \,F. \,S\'anchez\inst{11},
  R. \,Somerville\inst{4}, A. \,Taylor\inst{12},
  L. \,Wisotzki\inst{10}, \and X. \,Zheng\inst{15} }

\offprints{Irina Marinova; \email{marinova@astro.as.utexas.edu}}

\institute{Department of Astronomy, University of Texas at Austin, Austin, TX, USA 
\and
  School of Physics and Astronomy, University of Nottingham,
  Nottingham, UK
\and
  Laboratoire d'Astrophysique, \'Ecole Polytechnique F\'ed\'rale de
  Lausanne, Sauverny, Switzerland
\and
  Max-Planck-Institut f\"{u}r Astronomie, Heidelberg, Germany
\and 
  National Research Council (NRC) -- Herzberg Institute of
  Astrophysics, Victoria, BC, Canada
\and
  Space Telescope Science Institute, Baltimore, MD, USA
\and
  Institute of Cosmology and Gravitation, University of
  Portsmouth, Portsmouth, UK
\and
  Department of Physics and Astronomy, University of Waterloo, Warterloo, 
  ON, Canada
\and
  Institute for Astro- and Particle Physics,
  University of Innsbruck, Innsbruck, Austria
\and
  Astrophysikalisches Insitut Potsdam, Potsdam, Germany
\and
 Centro Astron\'omico Hispano-Alem\'an, Calar Alto, Spain
\and
  The Scottish Universities Physics Alliance, University of Edinburgh, 
  Edinburgh, UK
\and
  Department of Astronomy, University of Massachusetts, Amherst, MA, USA
\and
  Sub-department of Astrophysics, University of Oxford, Oxford, UK
\and
  Purple Mountain Observatory, Chinese Academy of Sciences, Nanjing, China
\and
  University of Texas, McDonald Observatory, Fort Davis, TX, USA
\and
  Department of Physics, University of Missouri, Kansas City, MO, USA
}

\authorrunning{Marinova et al.}

\titlerunning{Barred disks in dense environments}

\abstract{We investigate the properties of bright ($M_{V} \le 
 -18$) barred and unbarred disks in the Abell 901/902 cluster system
 at $z\sim$~0.165 with the STAGES HST ACS survey. To identify and
 characterize bars, we use ellipse-fitting. We use visual
 classification, a S{\'e}rsic cut, and a color cut to select disk
 galaxies, and find that the latter two methods miss 31\% and 51\%,
 respectively of disk galaxies identified through visual
 classification.  This underscores the importance of carefully
 selecting the disk sample in cluster environments.  However, we find
 that the global optical bar fraction in the clusters is $\sim$~30\%
 regardless of the method of disk selection. We study the relationship
 of the optical bar fraction to host galaxy properties, and find that
 the optical bar fraction depends strongly on the luminosity of the
 galaxy and whether it hosts a prominent bulge or is bulgeless.
 Within a given absolute magnitude bin, the optical bar fraction
 increases for galaxies with no significant bulge component. Within
 each morphological type bin, the optical bar fraction increases for
 brighter galaxies.  We find no strong trend (variations larger than a
 factor of 1.3) for the optical bar fraction with local density within
 the cluster between the core and virial radius ($R\sim$~0.25 to 1.2
 Mpc).  We discuss the implications of our results for the evolution
 of bars and disks in dense environments.

}

\maketitle{}

\section{Introduction}
\label{intro}

Mounting evidence suggests that a dominant fraction of bulges in
fairly massive ($M_{\ast}\sim 10^{10} - 10^{11}$~M$_{\odot}$)
$z\sim$~0 galaxies (e.g., Weinzirl et al. 2009) as well as the bulk of
the cosmic star formation rate density since $z < $~1 in intermediate
and high mass galaxies, are not triggered by ongoing major mergers
\citep{Hammer05, Wolf05, Bell05, JogeeIAU, Jogee08} but are likely
related to a combination of minor mergers and internally-driven
secular processes

Barred disks have been studied extensively in the nearby
universe. Quantitative methods show that approximately 45\% of disk
galaxies are barred in the optical at $z\sim$~0 \citep{MJ07, Reese07,
  BJM08, Aguerri08}.  Bars drive gas to the centers of galaxies,
where powerful starbursts can be ignited \citep{Schwarz81, KK04,
  Jogee05, Sheth05}. These starbursts caused by bar-driven inflow may
help build up central, high $v/\sigma$, stellar concentrations called
disky bulges or pseudobulges \citep{Kormendy82, Kormendy93, Jogee05,
  Weinzirl08}. Bars can also create boxy/peanut bulges through
vertical buckling \citep{Bureau99, Ath05, MVShlos06} and resonance
scattering \citep{Combes81, Combes90}.

But what is the relationship between bar-driven secular evolution and
environmental effects? Little is known about barred galaxies in dense
environments, as the situation is complicated by the relative
importance of processes such as ram pressure stripping, galaxy tidal
interactions, mergers, and galaxy harassment. In addition, further
complications are introduced by the fact that the bar fraction and
properties in clusters depend on the epoch of bar formation and the
evolutionary history of clusters. We explore these questions using the
Space Telescope Abell 901/902 Galaxy Evolution Survey (STAGES;
\citealp{Gray08}).

\section{Data and sample}
\label{data}

The main data for this study comes from the STAGES survey
$0\fdg5\times0\fdg5$ degree HST/ACS F606W mosaic of the Abell 901/902
cluster system at $z\sim$~0.165. Spectro-photometric redshifts for all
galaxies are available from the COMBO-17 survey \citep{Wolf04}, with
an accuracy of $\delta z/(1+z)\sim$~0.01 for the sample used in this
study. In addition, X-ray maps of the intra-cluster medium (ICM)
density from XMM-Newton, UV from GALEX, Spitzer 24$\mu$m coverage,
dark matter maps from weak lensing \citep{Heymans08}, and stellar
masses \citep{Borch06} are available for this field.

We use the spectro-photometric redshifts to select a sample of 785
bright ($M_{V} \le -18$), cluster galaxies. The point spread function
(PSF) for the ACS images is $0\farcs1$, corresponding to $\sim$~282 pc
at $z\sim$~0.165\footnote{We assume in this paper a flat cosmology
  with $\Omega_{\rm m} = 1 - \Omega_{\Lambda} = 0.3$ and $H_{\rm 0}$
  =70~km~s$^{-1}$~Mpc$^{-1}$.}. While this resolution is sufficient
for detecting primary bars in spiral galaxies, bars in dwarf galaxies
may often be smaller than the detection limit
(Sect.~\ref{method}). Therefore, in this study we focus on galaxies
brighter than $M_{V} = -18$ to avoid dwarf galaxies, to keep field
contamination low ($\sim$~10\%; \citealp{WGM05}), and to maintain
adequate resolution for bar detection through ellipse-fitting.  From
the 785 bright cluster galaxies we are able to ellipse-fit 762 (97\%;
Sect.~\ref{barmeth}).  This is the total sample from which all other
subsamples are derived in the subsequent analysis.

\section{Method}
\label{method}

\subsection{Selecting disk galaxies}
\label{diskselmeth}

Because bars are a disk phenomenon, the bar fraction $f_{\rm bar}$ is
defined as the number of barred disk galaxies over the total number of
disk galaxies:

\begin{equation}
f_{\rm bar} = \frac{N_{\rm barred}}{N_{\rm disk}} = \frac{N_{\rm
      barred}}{N_{\rm barred} + N_{\rm unbarred}}. 
\end{equation}
For this reason, it is necessary to select a sample of disk galaxies
from our total cluster sample.

We compare three widely used methods of selecting disk galaxies:
visual classification, $U-V$ blue-cloud cut, and S{\'e}rsic cut. We
visually classify all galaxies in the sample into broad bins of $B/T$:
`pure bulge', `bulge+disk', and `pure disk'. For the blue-cloud cut,
only galaxies on the blue cloud in color-magnitude space are selected
as disks. For the S{\'e}rsic cut, only galaxies with single-component
S{\'e}rsic index $n < 2.5$ are selected as disks.

\subsection{Identifying and characterizing bars}
\label{barmeth}

We use ellipse-fitting to identify and characterize bars (e.g.,
\citealp{Wozniak95, Friedli96, Jogee04, Knapen00, Sheth03, MJ07}). We
use the standard IRAF task `ellipse' to fit isophotes to each galaxy
out to a radius $a_{\rm max}$ where the surface brightness reaches sky
level. We generate radial profiles of the surface brightness,
ellipticity ($e$), and position angle (PA), as well as overlays of the
fitted ellipses onto the galaxy images. We use the radial profiles and
overlays (as an extra check) to classify galaxies as highly inclined
($i > 60^{\circ}$), unbarred, or barred. A galaxy is classified as
barred if (a)~the $e$ profile rises to a global maximum ($> 0.25$)
while the PA stays constant and (b)~after the global maximum near the
bar end, $e$ drops and the PA changes indicating the outer disk
region.  We do not attempt to classify highly inclined ($i >
60^{\circ}$) galaxies as barred or unbarred with
ellipse-fitting. However, morphological features such as bars and
spiral arms are classified regardless of inclination during visual
classification of the sample.

After excluding highly inclined galaxies (226 or 37\%) and those with
poor fits (32 or 5\%), we are left with 350 moderately inclined ($i <
60^{\circ}$), bright ($M_{V} \le -18$), ellipse-fitted,
visually-identified disk galaxies. To ensure adequate resolution for
bar identification, we make a cut in disk size $a_{\rm disk}$ of
3~kpc. This reduces our sample by 10 galaxies, leaving us with 340
bright, moderately inclined, ellipse-fitted, visually-identified disk
galaxies.

\section{Results}
\subsection{Disk selection}

\begin{figure*}[t!]
\resizebox{\hsize}{!}{\includegraphics[clip=true]{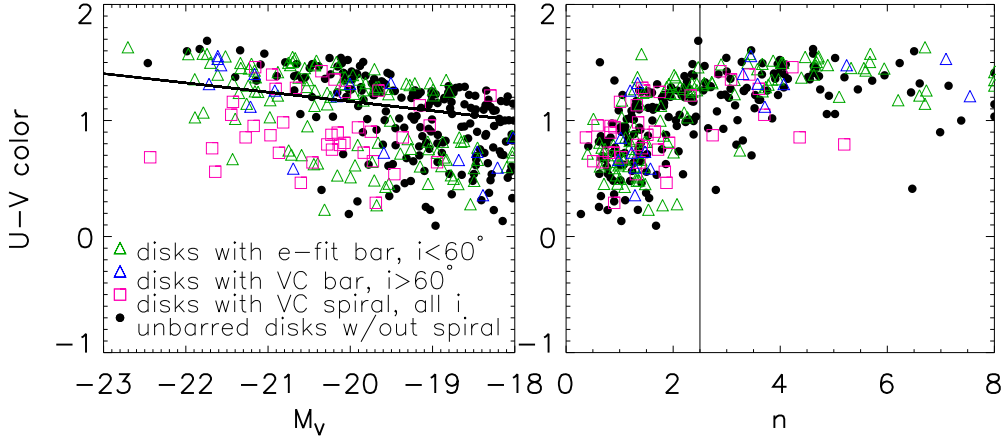}}
\vskip -2.8 in
\caption{\footnotesize
  (a)~Rest-frame $U-V$ color vs. $M_{V}$ absolute magnitude. 
  The solid line shows the separation between the red sample and blue
  cloud. (b)~Rest-frame $U-V$ color vs. S{\'e}rsic index $n$. The
  solid line shows the cut at $n = 2.5$. In both panels,
  moderately-inclined disks with ellipse-fit identified bars are shown
  as green triangles, while for inclined galaxies ($i>60^{\circ}$),
  barred galaxies identified through visual classification are plotted
  as blue triangles. Unbarred disks with spiral arms (from visual
  classification) are shown as pink squares for all inclinations. Disk
  galaxies identified with visual classification. Unbarred disks
  without spiral arms are plotted as black points. A large fraction of
  visually-identified disk galaxies are missed by the color cut and
  S{\'e}rsic methods.}
\label{diskmiss}
\end{figure*}

We compare three widely used methods of disk selection: S{\'e}rsic
cut, blue-cloud cut, and visual classification. The three methods pick
out 467, 341, and 608 ellipse-fitted disk galaxies,
respectively. Fig.~\ref{diskmiss}a shows that when a S{\'e}rsic cut is
used, 31\% of disk galaxies are missed compared to visual
classification. This underscores the difficulty in selecting disks
using S{\'e}rsic and color cuts, because of the prevalence of S0s and
red spirals in low-z cluster environments. When a blue-cloud cut is
used (Fig.~\ref{diskmiss}b), 51\% of visually-identified disk galaxies
are missed. There is inherent difficulty in distinguishing unbarred
disk galaxies without spiral arms from spheroids using visual
classification. We can obtain a lower limit to how many galaxies are
disks in the red sample or with $n > 2.5$, by only considering
galaxies exhibiting unambiguous disk signatures such as bars and
spiral arms. We find that at least 25\% (67/267) of galaxies with
$n>2.5$ and 22\% (84/390) of galaxies on the red sample exhibit such
features.

\subsection{Optical bar fraction}

For moderately inclined disks, the three methods of disk selection
yield a similar \textit{global} optical bar fraction ($f_{\rm
  bar-opt}$) of 34\%$^{+10\%}_{-3\%}$, 31\%$^{+10\%}_{-3\%}$, and
30\%$^{+10\%}_{-3\%}$, respectively. However, the \textit{global}
optical bar fraction may not reveal the full picture, as there is
evidence that in the field, the optical bar fraction is a function of
$B/T$, and possibly color \citep{BJM08, Aguerri08}. We therefore
explore how the optical bar fraction varies as a function of $M_{V}$
and qualitative measures of $B/T$ in the Abell 901/902 cluster
system. We note that while we did not quantitatively measure $B/T$ in
the sample, we visually classified the extreme ends of the
distribution into the broad classes described
above. Table~\ref{fbar_BT} shows that in the cluster, the optical bar
fraction is a strong function of $B/T$ and absolute magnitude. At a
given morphological type, $f_{\rm bar-opt}$ rises for brighter
galaxies. At a given $M_{V}$, $f_{\rm bar-opt}$ rises for `pure disk'
galaxies.

\begin{table*}
\caption{Optical bar fraction as a function of host absolute
magnitude and morphological class}
\label{fbar_BT}
\begin{center}
\begin{tabular}{lcc}
\hline
\\
$M_{V}$ range & Bulge+Disk & Pure Disk\\
\hline
\\
$-18 \ge M_{V} > -19$& 13\%$\pm4\%$ (8/63) &  39\%$\pm8\%$ (16/41) \\
$-19 \ge M_{V} > -20$& 19\%$\pm4\%$ (17/88)&  59\%$\pm9\%$ (16/27) \\
$-20 \ge M_{V} > -21$& 40\%$\pm6\%$ (29/73) & 63\%$\pm17\%$ (5/8)  \\
$-21 \ge M_{V} > -22$& 63\%$\pm8\%$ (22/35) &  ---  \\
\hline
\end{tabular}
\end{center}
\end{table*}

\begin{figure*}[t!]
\vskip -1 in
\resizebox{\hsize}{!}{\includegraphics[clip=true]{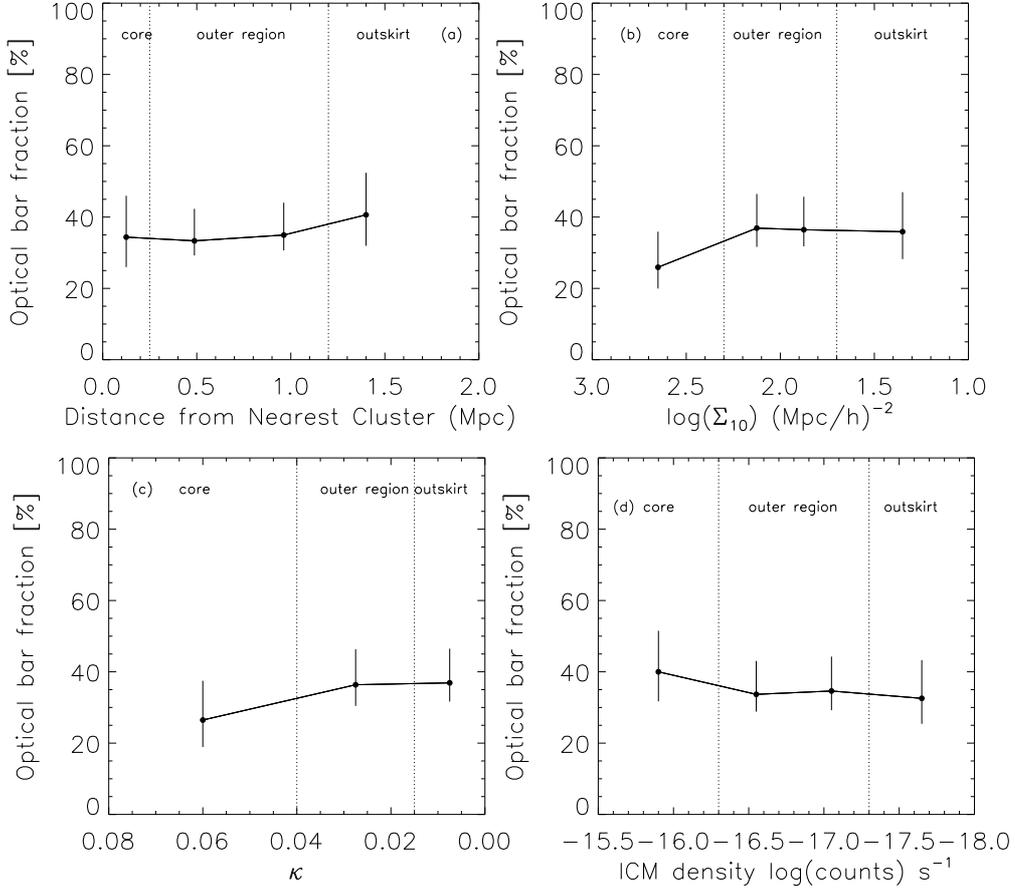}}
\vskip -0.7 in
\caption{\footnotesize
  The fraction of barred galaxies a function of:~ (a) distance from
  nearest cluster center, (b) log$\Sigma_{10}$, (c) $\kappa$, and (d)
  ICM density.  Bar classifications are from ellipse fits and disks
  are identified by visual classification. The vertical dashed lines
  denote the core radius at 0.25~Mpc and the virial radius at
  1.2~Mpc. Between the core and the virial radius of the cluster
  $f_{\rm bar-opt}$ varies at most by a factor of $\sim$~1.3, allowed
  by the error bars.}
\label{densityall}
\end{figure*}

The result that $f_{\rm bar-opt}$ rises toward disk-dominated galaxies
may be interpreted in the context of bar formation models. If bars
form and grow through the swing amplification, a large bulge that
leads to an inner Lindblad resonance (ILR) may inhibit bar formation
by cutting off the feedback loop. Early simulations with rigid halos
suggested that a hot DM halo component can inhibit the formation of a
bar (e.g., \citealp{OP73}). However, more recent simulations with live
halos show that bars grow through the exchange of angular momentum
with the DM halo \citep{DebSell98, DebSell00, Ath03}. These studies
suggest that the evolution of the bar is intimately tied to the mass
distribution of the DM halo within the radius of the galaxy disk. If
fainter galaxies have more dominant DM halos (e.g., \citealp{Persic96,
  Kassin06}), bars may develop and grow more slowly in these galaxies,
however in the end, they may be stronger.

\subsection{Bars as a function of local density}
   
How does the optical bar fraction vary as a function of local density
in the cluster? We use four available measures of local density:
projected distance to nearest cluster center, $\kappa$, $\Sigma_{10}$,
and ICM density. $\kappa$ is line-of-sight projected surface mass
density, and $\Sigma_{10}$ is the local galaxy number
density. Figure~\ref{densityall} shows that the optical bar fraction
varies at most by a factor of 1.3 between the core and virial radius
of the clusters.  Our results agree with Aguerri et al. (2009) who
find no dependence of the optical bar fraction on local environment
density, over a wide range of environments comparable to the outer
region and outskirt of our cluster sample. Theoretical models of bar
formation in a cosmological context, suggest that interaction with the
halo sub-structure induces bars \citep{Shlos08}.  Because this
substructure is present in all environments, these models imply a
similar bar fraction across a large range of environment densities,
which is consistent with our results.

In the core, our data does not allow a firm conclusion, because the
four tracers of local density show opposing trends, projection effects
become severe, and number statistics are low. Previous studies using
visual identification of bars have argued that the optical bar
fraction increases in the Coma and Virgo cluster cores
\citep{Thompson81, Andersen96}. If this trend is real, it could be
caused by competing effects. Our results suggest that because
bulge-dominated galaxies, which are prevalent in cluster cores, host a
lower fraction of bars, the bar fraction should decrease
there. However, the high velocity dispersion in the core regions
favors numerous, short-lived, tidal interactions, which may increase
the bar fraction, while preserving intact the galaxy disk.

\begin{acknowledgements}
IM, SJ, and AH acknowledge support from NSF grant AST 06-07748, NASA
LTSA grant NAG5-13063, as well as HST G0-10395. 
\end{acknowledgements}

\bibliographystyle{aa}

\end{document}